\documentclass[10pt]{article}
\usepackage{epsfig}
\usepackage{amsfonts}
\usepackage{graphicx}
\usepackage{epsfig} 

\date{\today}

\begin{document}

\title{Skyrmion and Skyrme-Black holes in de Sitter spacetime}
\author{{\large Yves Brihaye \footnote{yves.brihaye@umh.ac.be}
\ \ and \ \ {Terence Delsate}}\\
\small{
Facult\'e des Sciences, Universit\'e de Mons-Hainaut,
B-7000 Mons, Belgium }\\
{ }\\
}
\date{\today}

\maketitle
 
\begin{abstract} 
Numerical arguments are presented for 
the existence of regular and black hole
solutions of the Einstein-Skyrme equations
with a positive cosmological constant.
These classical configurations
approach asymptotically the de Sitter spacetime.
The main properties
of the solutions and the differences with respect to the
asymptotically flat ones are discussed. 
It particular our results suggest that, for a positive
cosmological constant, the mass evaluated as timelike
infinity in infinite.
Special emphasis
is set to De Sitter black holes Skyrmions which display
two horizons.
\end{abstract}

\section{Introduction}
It is almost two decades than the first examples of
hairy black holes are known \cite{luckock}. This early construction
of gravitating objects presenting both, an event horizon and a non trivial
structure of the matter fields, was first achieved in a context where the
matter fields are described in terms of a non-linear sigma-model,
for instance the Skyrme model \cite{skyrme}.
Initially, the Skyrme model was proposed more than forty years ago as an
effective model for chiral symmetry breakdown in quark models; 
the main fields are the {\it pion} particles and  soliton-like solutions
of the equations -the Skyrmions- are interpreted as the {\it nucleons};
see e.g. \cite{sutcliffe} for a complete and recent review of the topic.

Although the initial purpose of the Skyrme model was far from being
coupled to gravity, the classical equations resulting from its coupling
to gravity constitute a rich system of equations
where both (stable) gravitating solitons and 
hairy black holes solutions exist.
These were studied in great details in \cite{droz,bizon},
a recent review of these
solutions and motivations can be found in \cite{review}.
When space-time is imposed to be asymptotically
flat, the gravitating skyrmions exist in two branches indexed by
an effective coupling
constant $\alpha^2 \equiv 4 \pi G F_{\pi} $ where $G$ denotes Newton's
constant  and $F_{\pi}$, the pion decay constant,
is the coupling constant of the standard Skyrme model.
The two branches merge at a maximal
value, say $\alpha_{max}$. The solution with the lowest energy
smoothly approaches the
flat Skyrmion in the $\alpha \rightarrow 0$-limit and is known to be
stable on the basis of topological arguments.

Recently, the Einstein-Skyrme model was reconsidered by supplementing
the equations with a negative cosmological constant \cite{sawado}
and strong numerical evidence was given
that asymptotically anti-DeSitter hairy black hole Skyrmion exist as well.
More precisely,
the authors of Ref. \cite{sawado} have shown that
gravitating Skyrmion solution
exist with a metric approaching asymptotically 
the Anti-deSitter space for values of the cosmological constant
 $\vert \Lambda \vert$ lower than a maximal value, 
 say $\vert \Lambda \vert = \vert \Lambda_{max} \vert$.
Similarly to the case $\Lambda=0$, two branches
of solutions exist. When $\alpha$ is fixed and $\Lambda$ varies,
the two branches  terminate at the maximal value.
Another interesting issue of these calculations is
that not only the solutions
corresponding to the branch of lowest energy are stable. 
It seems that stable solutions
are available on the two branches.

Although there are many reasons to study AdS 
black holes and solutions with such an asymptotics
(see e.g. the AdS/CFT correspondance \cite{maldacena,witten}
and/or the brane world cosmology
arguments \cite{randall,kaloper}),
De Sitter (dS) space-time enjoyed recenty a huge interest in
theoretical physics for a
variety of reasons.
First at all, the observational evidence accumulated in the last
years (see, $e.g.$, ref.~\cite{data}), seems to favour the idea that the
physical universe has an accelerated expansion. The most common explanation
is that the expansion is driven by a small positive vacuum energy
($i.e.$ a cosmological constant $\Lambda>0$), implying
 spacetime to be asymptotically dS.
Furthermore, dS spacetime plays a central
role in the theory of inflation (the very rapid accelerated expansion in the
early universe), which is supposed to solve the cosmological flatness and
horizon puzzles.
Several results in the literature suggest that the conjectured
dS/CFT correspondence
has a number of similarities with the Anti-de Sitter/CFT
correspondence, although
many details and interpretations remain to be clarified
(see \cite{Klemm:2004mb} for a recent review and a large set
of refences on this problems).
In view of these developments, an examination of
the classical solutions of gravitating
fields in asymptotically  dS spacetimes seems appropriate.

Several solutions of this type were considered in the framework
of the Einstein-Yang-Mills equations \cite{Volkov:1996qj,Linden:2000ev,
Breitenlohner:2004fp} .
The  solutions of the Einstein-Maxwell theory  with $\Lambda>0$
have been discussed in a dS/CFT context in \cite{Astefanesei:2003gw},
multi-black hole configurations being considered as well.
In a recent paper \cite{bhr} the Einstein-Yang-Mills-Higgs
equations was considered
in the context of a positive cosmological constant.
One main result of this analysis is that
DeSitter monopole and sphaleron exist for small enough
value of the cosmological constant.
Again, they exist in two branches for $\Lambda \leq \Lambda_{max}$ but,
   computing the mass at timelike infinity
according to the formalism developped in
\cite{Gibbons:1977mu,Balasubramanian:2001nb,Balasubramanian:1999re}
we find that they do not have a finite mass as long as $\Lambda > 0$.

In this paper we  consider the Einstein-Skyrme equations for a positive
cosmological constant. We explore static, spherically symmetric
configurations of the metric and matter fields.
As a consequence of the
positive cosmological  constant $\Lambda$, these solutions
approach a DeSitter space-time asymptotically and present a
cosmological horizon. Because of the spherical symmetry, this
horizon occurs on a sphere, at an
finite value of the radial variable $r$.
We succeeded in solving the equations numerically in
both cases~: regular and black hole solutions.

The paper is organized as follows:
in Sect.2 we discuss the lagrangian, the
ansatz, the relevant boundary conditions
and establish the equations. In Sect.3 we
discuss the numerical solutions for both cases
(i) solutions regular at the origin and
(ii) solutions presenting an event horizon at some finite value $r=r_h$.
In both case the solutions present a cosmological horizon at $r=r_c$.
The behaviour of the fields in the interior of the event horizon is briefly
discussed as well.
\section{The Einstein-Skyrme Lagrangian}
\subsection{Action principle}
The action for a gravitating $SU(2)$ Skyrme model is
\begin{eqnarray}
\label{lag0}
{\cal S} =\int_{\mathcal{M}} d^4x \ \sqrt{-g}  \left( \frac{1}{16\pi G}
(R-2 \Lambda) + \mathcal{L}_{M}\right)
\end{eqnarray}
with  Newton's constant $G$ and  cosmological constant $\Lambda$. The 
matter part of the Lagrangian density chosen as the Skyrme model~:
\begin{eqnarray}
\label{lag0}
\mathcal{L}_{M}= 
\frac{F^2_{\pi}}{16} g^{\mu \nu} {\rm tr} (L_{\mu} L_{\nu})
+ \frac{1}{32 e^2}  g^{\mu \nu} g^{\rho \sigma}
{\rm tr} ([L_{\mu}, L_{\rho}] [ L_{\nu}, L_{\sigma} ])
\end{eqnarray}
The basic matter field,  denoted
$U(x)$, takes value in SU(2)
while the combination
$L_{\mu} \equiv U^{\dagger} \partial_{\mu} U$ has values in the
Lie algebra su(2). Here $F_{\pi},e$
represent the two coupling constants of the theory.
In the context of hadron physics,  $F_{\pi}$ is the pion
decay constant and $e$ is the Skyrme constant which ensures
the stability of the Skyrmion. Throughout the paper, we assume
the pion fields to be massless.
\subsection{Spherically symmetric ansatz}
In flat space, the fields equations associated with the
Skyrme model admit an extremely rich pattern of solutions
(see e.g. \cite{sutcliffe} for a recent review and
references therein).
However, here we will restrict to the spherically symmetric
solutions.
The spherically symmetric Skyrmion solution is constructed
by imposing the static, hedgehog ansatz for the chiral field~:
\begin{equation}
       U(x) = U(\vec r) = \cos f(r) + i \hat x \cdot \vec \tau \sin f(r)
\end{equation}
 where $\hat x \equiv \vec r / r$ and $\vec \tau$ denotes the Pauli
 matrices.

For the metric, we use the standard spherically symmetric
line element
\begin{equation}
\label{metric}
ds^{2}=\frac{dr^{2}}{N(r)}+r^2(d\theta^{2}+\sin^{2}\theta d\varphi^{2})-
\sigma^2(r)N(r)dt^{2}
\end{equation}
where we conveniently parametrize the
metric function $N(r)$ according to
\begin{equation}
N(r)=1-\frac{2m(r)}{r}- \frac{\Lambda r^2}{3}
\end{equation}

 The classical energy due to the matter fields can then be expressed
 in term of the following reduced functional
 \begin{equation}
    E_S = 4 \pi \frac{F_{\pi}}{e} \int
    ( \frac{1}{8} N u f'^2 +
\frac{v}{4 x^2 }) \sigma \ dx   \ \ \ , \ \ \   x \equiv F_{\pi} e r
 \end{equation}
with
\begin{equation}
       u \equiv x^2 + 8 \sin^2 f \ \ \ , \ \ \
       v \equiv \sin^2 f (x^2 + 2 \sin^2 f)
\end{equation}
Here a rescaled  radial coordinates $x$ is introduced. From now on,
the primes denote  derivatives with respect to $x$.  Accordingly,
it is convenient to define $\mu(x) = e F_{\pi} m(r)$ and
$\tilde \Lambda = \Lambda/ e^2 F_{\pi}^2$.

\subsection{Field equations}
The variational equations associated with the above functional
are called the Einstein-Skyrme equations.
Within the spherically symmetric ansatz, they reduce 
to the following system of three non-linear differential equations
\begin{eqnarray}
\label{e1}
\mu'&=& \frac{\alpha^2}{8} (N u f'^2 + \frac{2 v}{x^2})   \\
\label{e2}
\sigma'&=& \sigma \frac{\alpha^2}{4 x} u f'^2  \\
\label{e3}
(N \sigma  u f')'&=&
\sigma (4 N f'^2 + 1 + \frac{4 \sin^2 f}{x^2}) \sin 2 f  \\
\end{eqnarray}

\subsection{Boundary conditions}
We want the generic line element (\ref{metric}) to describe a nonsingular,
asymptotically de Sitter spacetime outside a cosmological horizon located 
at $x=x_c>0$.
In addition we require both possibilities of 
either a regular solution on the line
$[0,\infty]$ or an event horizon at $x = x_h \leq x_c$.
Here $N(x_h)=0$ and $N(x_c)=0$ are only coordinate singularities 
where all curvature
invariants are finite. Nonsingular extensions across these null
surfaces can be found.
The regularity assumption implies that all
curvature invariants at $x=x_c$ are finite.

The regularity conditions at $x=0$ are
\begin{equation}
    \mu(0) = 0 \ \ , \ \ f(0) =  \pi
\end{equation}

Examining closely the asymptotic values of the equations for $\Lambda >0$,
it turns out that the fields can take one of
the two following asymptotic forms
\begin{equation}
\label{asymp1}
   f = q +  \frac{1}{x^2} \frac{3}{2 \Lambda} \sin 2q + O(1/ x^3)
   \ \ , \ \
   m(r) =  \frac{\alpha^2}{4}  (\sin^2 q) x + M + O(1/x) \ \ , \ \
   \sigma(x) = 1 - \frac{\alpha^2}{4} \frac{c^2}{x^4} + O(1/x^5)
\end{equation}
 here $q$ denotes an arbitrary constant; or
 \begin{equation}
\label{asymp2}
 f = \frac{F}{x^3} + O(1/x^5) \ \ , \ \
 m = M + \frac{\alpha^2 \Lambda}{8} \frac{F^2}{x^3} + O(1/x^5) \ \ , \ \
 \sigma = 1 - \frac{3 \alpha^2}{8} \frac{F^2}{x^6} + O(1/x^7)
\end{equation}
 where $F$ is a constant.
The form (\ref{asymp2}) leads to a finite mass $M$ for the solution;
for $\Lambda < 0$, it is precisely the form obeyed by the
solutions constructed in \cite{sawado}.

At the event or cosmological horizon,
(i.e. a value corresponding to a zero of the metric function $N(r)$)
the regularity of the equation of the chiral
field leads to the following condition
\begin{equation}
\label{horizoncond}
    (uf'(x(\Lambda x^2 + \alpha^2 \frac{v}{x^2}) - x) + 
    2 \sin(2f) (x^2 + 4 \sin^2f) \vert_{x=x_h \ {\rm or} \  x=x_c} = 0
\end{equation}

The numerical integration is  first performed  on
$[0,x_c]$
(or on $[x_h,x_c]$ in the case of black holes)
with the condition (\ref{horizoncond}) imposed as a boundary
condition at $x=x_c$ (or at $x=x_h$ and $x=x_c$ in case of black hole).
 Then the solution
is  continued by integrating
on $[x_c,\infty]$, again, imposing the condition
(\ref{horizoncond}) at $x=x_c$. The asymptotic behaviour (\ref{asymp1})
or (\ref{asymp2}) will be determined by means of this second integration,
 together with the corresponding value
 of $q$ or $F$ .

Both the event  and the cosmological
horizon have their own  surface gravity $\kappa$
given by
\begin{eqnarray}
\nonumber
\kappa^2_{h,c}=- \frac{1}{4}g^{tt}g^{rr}
(\partial_r g_{tt})^2\Big|_{r=r_h,r_c}~~,
\end{eqnarray} 
the associated Hawking temperature being $T_H=|\kappa|/(2\pi)$.
\subsection{Known solutions}
For several limits on the different parameters,
the solutions of the above equations are well known.
The  Schwarzschild-de Sitter solution corresponds to
\begin{eqnarray}
\label{SdS}
f(r)= k \pi ,~~~~~\sigma(r)=1,~~~~
N(r)=1-\frac{2M}{r}-\frac{\Lambda r^2}{3},
\end{eqnarray}
and describes a black hole inside a cosmological
horizon as long as $N(r)$ has two positive zeros, $i.e.$
$M< 1/3 \sqrt{\Lambda}$.

In the flat limit $\alpha = \Lambda = 0$, the Skyrmion
solution is recovered.
For $\Lambda=0$, $\alpha \neq 0$ the gravitating Skyrmion
(resp. Skyrme-black holes) are obtained.

\section{Numerical results}
The system of equations depends on three parameters
$\Lambda$, $\alpha$ and  $x_h$.
The case $\Lambda > 0$ leads to the occurence of a cosmological
horizon at $x = x_c$ with $N(x_c) = 0$.
To integrate the equations, we used the differential
equation solver COLSYS which involves a Newton-Raphson method
\cite{COLSYS}. The equations are first solved on the interval
 $[0,x_c]$ for regular solutions and on the interval
$[x_h,x_c]$ for black holes where $x_h$ corresponds to the
event horizon.  From this, the value of $\Lambda$
can be determined numerically. From the data of the numerical profiles obtained
on this finite interval, we are able to extend the solution outside the
cosmological horizon i.e. for $x \in [x_c,\infty]$. 
In the case where an event horizon is present, we further integrated
inside the event horizon, i.e. for $x \in [\epsilon,x_h]$;
here $\epsilon$  denotes a small cut off. Since the origin constitutes
an essential singularity of the metric ($N(0)= -\infty$),
the integration cannot be performed at the origin.
We will discuss this case in a special section. The different solutions are
constructed numerically with a absolute error lower than $10^{-6}$.
\subsection{Skyrme solutions in a fixed dS background}
If we set $\alpha=0$, we have trivially $\sigma=1$ and $N$ corresponds to the
dS solution in the vacuum. The matter equation then leads to the Skyrme
equation in the background of a deSitter space-time.
One solution of this type is represented
in Fig. 1. for $x_c=44.0$, corresponding to $\Lambda \approx 0.0031$.
Many features of these solutions are recovered in the presence of gravity
and will be discussed at length  in the next sections.
\subsection{Gravitating DeSitter Skyrmion}
We now set $\alpha > 0$.
If  regular conditions are imposed at the origin, the
system can be integrated first on $x \in [0, x_c]$, fixing $x_c$ by hand
imposing (\ref{horizoncond}) at $x=x_c$.
The corresponding value of $\Lambda$ is then determined numerically.
While $\alpha$ increases, we observe that the function $N(x)$
develops a minimum at $x = x_m$ with  $x_h < x_m < x_c$,
as illustrated on Fig. 1 for $\alpha = 0.3$.

In the case $\Lambda \ll 1$, the function $\mu$ seems to
attains a constant asymptotic value inside the sphere $x=x_c$, 
accordingly the parameter $M$ of Eq.(\ref{asymp1})
 can be determined directly. However, the cosmological
horizon decreases while  $\Lambda$ increases,  as a consequence the occurence
of $x_c$ outside the core of the Skyrmion does not persist for
large values of $\Lambda$ (typically for $\Lambda > 0.01$) and then
the integration of the equation for $r \in [x_c,\infty]$ turns out to be
necessary to refine the evaluation of the parameter $M$.
This can be achieved by using the data
of numerical the solution at $x=x_c$.

However, this is not the end of the story.
Indeed, the integration on $[x_c,\infty]$, reveals that the solutions
do not fall asymptotically on
the configuration (\ref{asymp2}) but rather on (\ref{asymp1})
where the value $q$ depends
non trivially on $\Lambda$. The occurence of the cosmological
constant therefore
prevents the chiral field to reach the zero value $f=0$ asymptotically.
Our numerical analysis strongly suggests that no solutions
decaying according to (\ref{asymp2}) exist.  An analytical
argument would however be necessary to state this result.

The parameter $q$ typically depends on $\Lambda$ and is determined
numericaly (typically, $q \approx 0.05$ for the values
corresponding to Fig. 2).
The numerical evaluation clearly suggests  that
$lim_{\Lambda \to 0} q / \Lambda $ is finite.
Accordingly the standard
 Skyrmion decay i.e. $f(r) \sim C/r^2$,   is recovered in the
gravitating but asymptotically flat limit
(note: we do not include any mass term for the chiral field
in this paper).  This result deeply contrasts with the case
$\Lambda < 0$ (see \cite{sawado})
which has no cosmological horizon and where
the asymptotic form (\ref{asymp2}) is obtained by a direct
integration on $[0,\infty]$. In fact we were able to construct 
numerically the
anti-de Sitter counterpart of our solution (i.e. for $\Lambda < 0$
and decay of type (\ref{asymp1})). These solutions do not have a finite
mass and were not emphasized in \cite{sawado}.

The physical consequences of this result are important.
Indeed, because of (\ref{asymp1}) the function $m(r)$
acquires a small linear dependance and does not stay constant asymptotically,
preventing the mass to stay finite. [We do not illustrate
this on a graphic but we refer to Fig. 2 where an (identical) phenomenon
is present in the asymptotic behaviour of a black hole.]
After an appropriate redefinition  of the radial variable $x$
and by using a standard argument, it can be shown that this property
of the mass leads asymptotically to a locally deSitter space-time
with an angular deficit given by  $4 \pi (1 - \alpha^2 (\sin^2 q)/2)$.

It is worth to point out that the feature of non finiteness of the
energy of soliton in asymptotically  deSitter space-time
 was already observed in \cite{bhr}; this was in the context of
spontaneously broken SU(2)-gauge fields theories, respectively with the
magnetic monopole (case of a Higgs triplet) and for sphaleron (case of a Higgs
doublet). 
Global monopoles are also studied in space-times involving
a cosmological constant \cite{bhr1}, in this case also the mass
function increases linearly and leads to an angular deficit.
This property persists in the presence of a cosmological constant.
In the present case, however, the mass evaluated at the cosmological
horizon is finite (see \cite{bhr}) are references therein.

The novel feature present in the case of the Skyrme field is
that the cosmological constant drives the radial function $f(r)$ away from its
standard asymptotic value $f(r)=0$. As a consequence the  chiral field
$U(\vec r)$ does not approach $U= 1_2$ asymptotically.

In flat space, the Skyrme solitons are largely characterized by their
baryon number.
This charge is defined as the integral
of the zero-component $B^0$ of the topological current
$B^{\mu}$~:
\begin{equation}
B \equiv \int \sqrt{-g} B^0 d^3 x \ \ , \ \
B^{\mu} = - \frac{1}{24 \pi^2} 
\varepsilon^{\mu \nu \rho \sigma} \frac{1}{\sqrt{-g}}
{\rm Tr} (L_{\nu}  L_{\rho} L_{\sigma})
\end{equation}
over a time-fixed section of space-time. It has an integer value
and is interpreted as the baryon number of the solution.

When the Skyrmion is considered in a asymptotically flat space-time
\cite{bizon} the charge $B$ is still an integer.
In the present context however, $B$  
stops to be an integer for several reasons.
In principle, we have to limit the integral defining $B$ to the 
domain $ 0 \leq r \leq r_c$ which corresponds to the limit of the observable
universe inside the cosmological horizon. The value of $B$ then deviates
from an integer because in general we have $f(r_c) \neq 0$.
However, even if we take advantage of the continuation  of the solution
for $r \in [r_c, \infty]$ to extend the space maximally, we find
after some algebra (see \cite{bgrt,sawado})
\begin{equation}
 B   = \frac{1}{2 \pi}(2 f - \sin 2 f)_{r=0} -
 \frac{1}{2 \pi}(2 f - \sin 2 f)_{r=\infty}
 = 1  - \frac{1}{2 \pi}(2q- \sin (2q))
\end{equation}
which is obviously not an integer.

Different mechanisms leading to non conservation of the baryon
number were constructed and examined e.g. in \cite{bgrt}. Our analysis
just reveals that supplementing  the Skyrme model with a positive cosmological
constant leads to the same feature.

\subsection{Black hole solutions}
When the conditions of an event horizon are
imposed at $r=r_h$ and for $\Lambda > 0$, the function $N$ possesses
two zeros. The system is solved first on the interval $[x_h, x_c]$
(we assume $r_h \ll r_c$ throughout all simulations) 
Again, the corresponding value of $\Lambda$
is determined numerically.
On Fig. 2 we present the profiles of the solutions for
$\alpha=0.3$, $r_h = 0.1$ and $x_c = 10$, corresponding to 
$\Lambda \approx 0.04$. On this figure, we clearly see that the
chiral function $f(r)$ does not approach zero asymptotically
and that the mass function $\mu$ starts increasing after it stays on
a plateau in the region of the cosmological horizon. We insist that 
this is in full aggrement with  (\ref{asymp1}).

While increasing the value of the cosmological constant
the numerical analysis shows that the solution (in fact black holes
and regular at the origin) exist up to some maximal value of $\Lambda$,
say up to $\Lambda = \Lambda_{max}$.
No solution seems to exist for  $\Lambda > \Lambda_{max}$ but
a second family of solution exist for  $\Lambda < \Lambda_{max}$.
For a given value of $\Lambda$ the mass inside associated with the
solutions on the second branch is greater than the corresponding mass
for the first branch.

Some physically relevant quantities
characterizing the solutions are presented in Fig. 3 in functions
of the cosmological constant parameter.
Here we plot namely the mass, the value of the cosmological horizon
and the values of the temperature at two horizons. The temperatures
at the cosmological horizon corresponding to the two branches are
the same, contrasting with the temperature at the event horizon
which comes out to be larger for the solutions on the second branch.

Our numerical results further indicate that, for fixed values of 
$\alpha, \Lambda$, the value  of
the parameter $q$ is larger on the second branch than on the first
(or main) one.
Not that the construction of the second branch becomes rather difficult
when reaching small values of $\Lambda$. That's why the second branch on Fig.
3 seems incomplete but we believe that the second branch extend backward
to $\Lambda = 0$ and we plan to solve this numerical difficulties in near
future.
\subsection{Inside the event horizon}
The question of integrating an hairy black hole solution
inside the event horizon
was adressed in \cite{blm1} for Einstein-Yang-Mills (EYM) black holes
with and without a Higgs field.
The authors pointed out serious numerical
difficulties that are met
when the integration inward the horizon is performed by using
the numerical data
available from the integration in $x \in [x_h,\infty]$. They called
the different phenomenon attached to the interior solution "mass inflation
inside hairy black holes".
When we attempt to integrate the Einsten-Skyrme equation  
for $x < x_h$ by using the  data available from the
integration on $x\in [x_h,x_c]$
we are immediately faced  numerical difficulties which, likely,
have the same origin than in \cite{blm1}.
Similarly to the case of EYM and EYMH equations, we notice
the occurence of regions inside the event horizon
where the derivative of the function $f$ varies suddently.
This is illustrated on Fig. 4. In addition, the chiral function $f(x)$
seems to deviate from the standard value $f(0)=\pi$ and to approach
a different value in the limit $x \to 0$.
In the case of EYM, it was observed, similarly,
that the gauge function $w(r)$ does not approach $w=1$ when $r \to 0$.
A more detailed analysis of this part of the solution is 
under investigation.
\section{Conclusions}
This work was partially motivated by the question on how a positive
cosmological constant will affect the properties of a gravitating skyrmion.
To the best of our knowledge, this question has not yet been addressed in the 
literature.
The unexpected result of our analysis is, in our opinion, the
fact that the presence of a positive cosmological constant prevents
the skyrmion to have an integer topological number. The physical
consequence of this result is that the solutions do not have a 
finite mass evaluated at timelike infinity.
This suggests
that, in the background of a varying cosmological constant, e.g.
during inflation, the baryon number of the system could be violated.
The analysis of the solutions reported here is minimal and will
be extended in near future.
\\
\\
\medskip
\noindent
{\bf\large Acknowledgements} \\
YB gratefully acknowledges E. Radu for numerous discussions
and the Belgian F.N.R.S. for financial support.


\newpage
\begin{figure}
\epsfysize=22cm
\epsffile{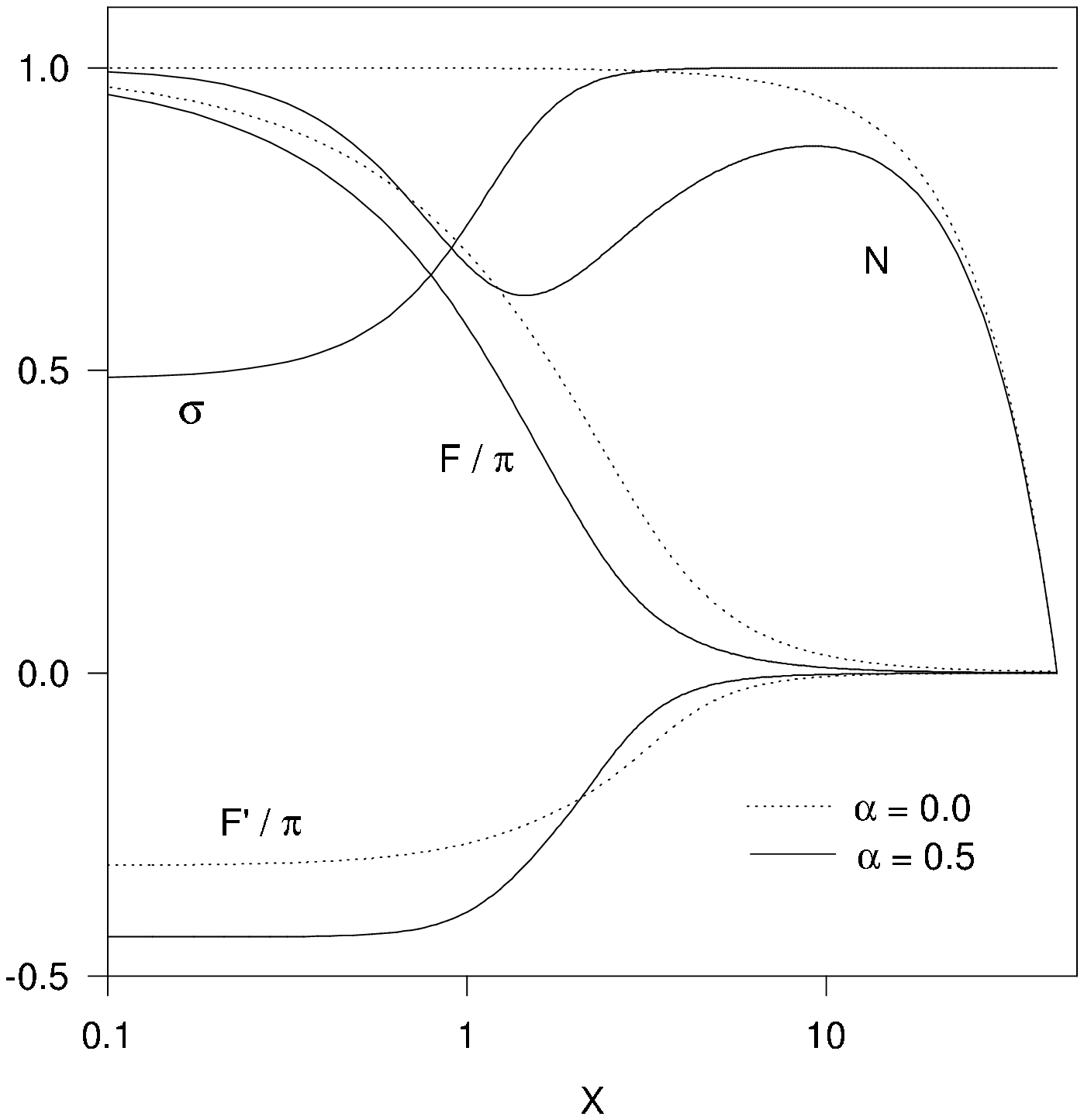}
\vskip -3cm
\caption{\label{fig1}
The profile for the DeSitter-Skyrme soliton for $\alpha=0.0$
and $\alpha=0.3$
}
\end{figure}

\newpage
\begin{figure}
\epsfysize=22cm
\epsffile{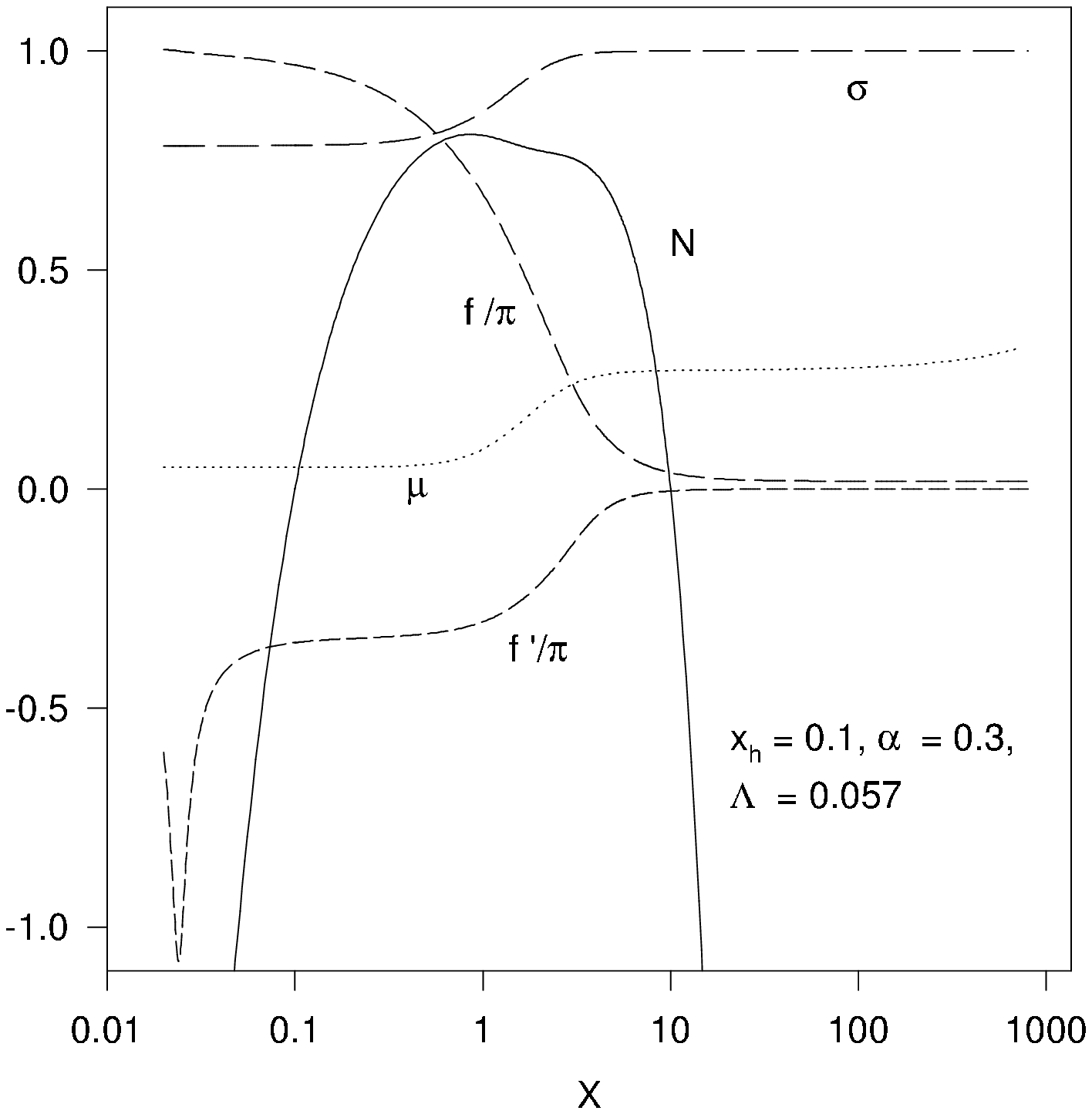}
\vskip -3cm
\caption{\label{fig2}
The profile for the DeSitter-Skyrme black hole for $\alpha=0.3, x_h=0.1$
}
\end{figure}

\newpage
\begin{figure}
\epsfysize=22cm
\epsffile{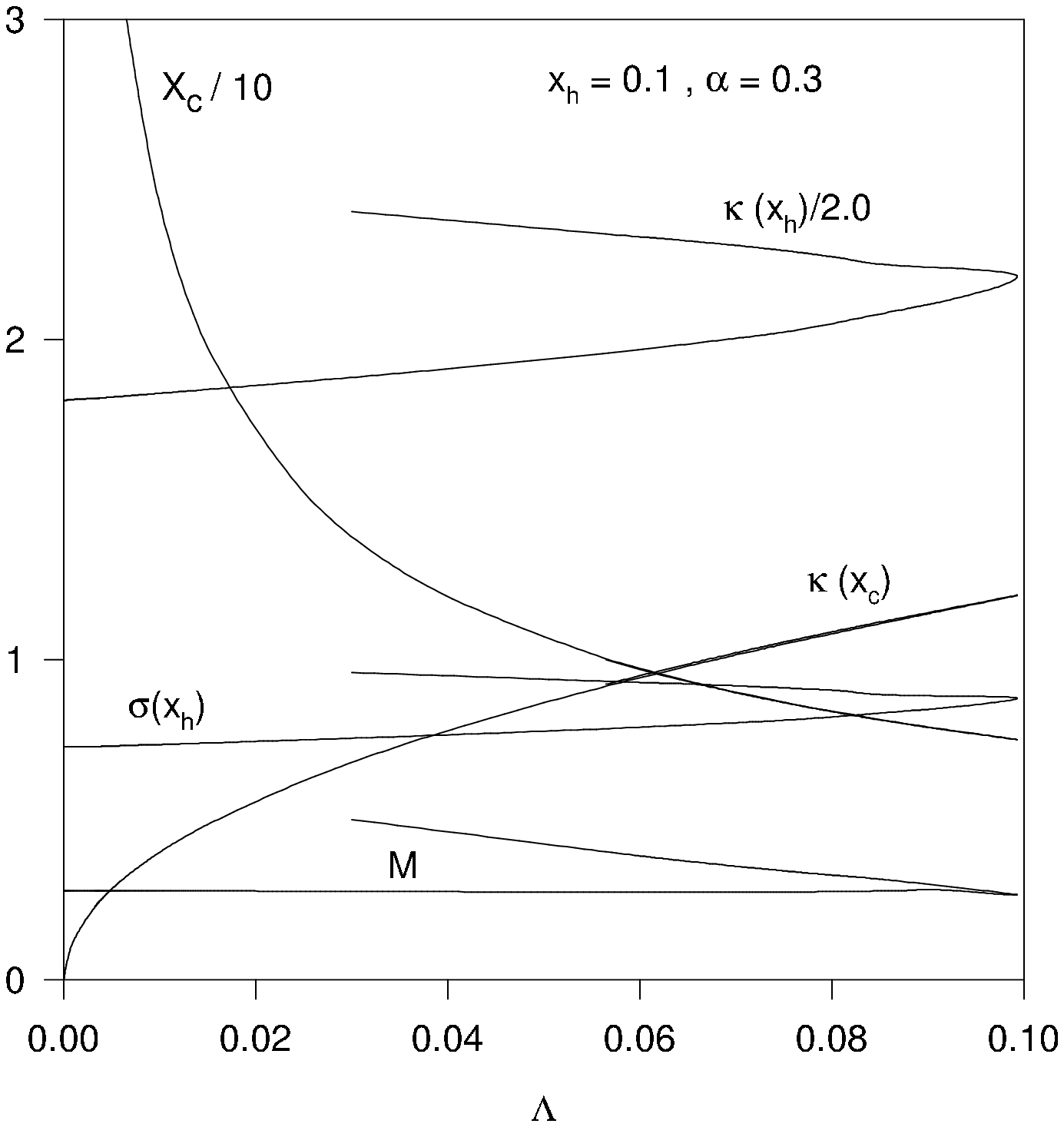}
\vskip -3cm
\caption{\label{fig3} The value of the cosmological horizon
$x_c$, the value of the metric function $\sigma$ at the event
horizon,
 the mass of the solution of the DeSitter Black holes 
 and the temperatures at the two horizons are
given as functions of $\Lambda$ for $\alpha = 0.3$, $x_h=0.1$.
}
\end{figure}

\newpage
\begin{figure}
\epsfysize=22cm
\epsffile{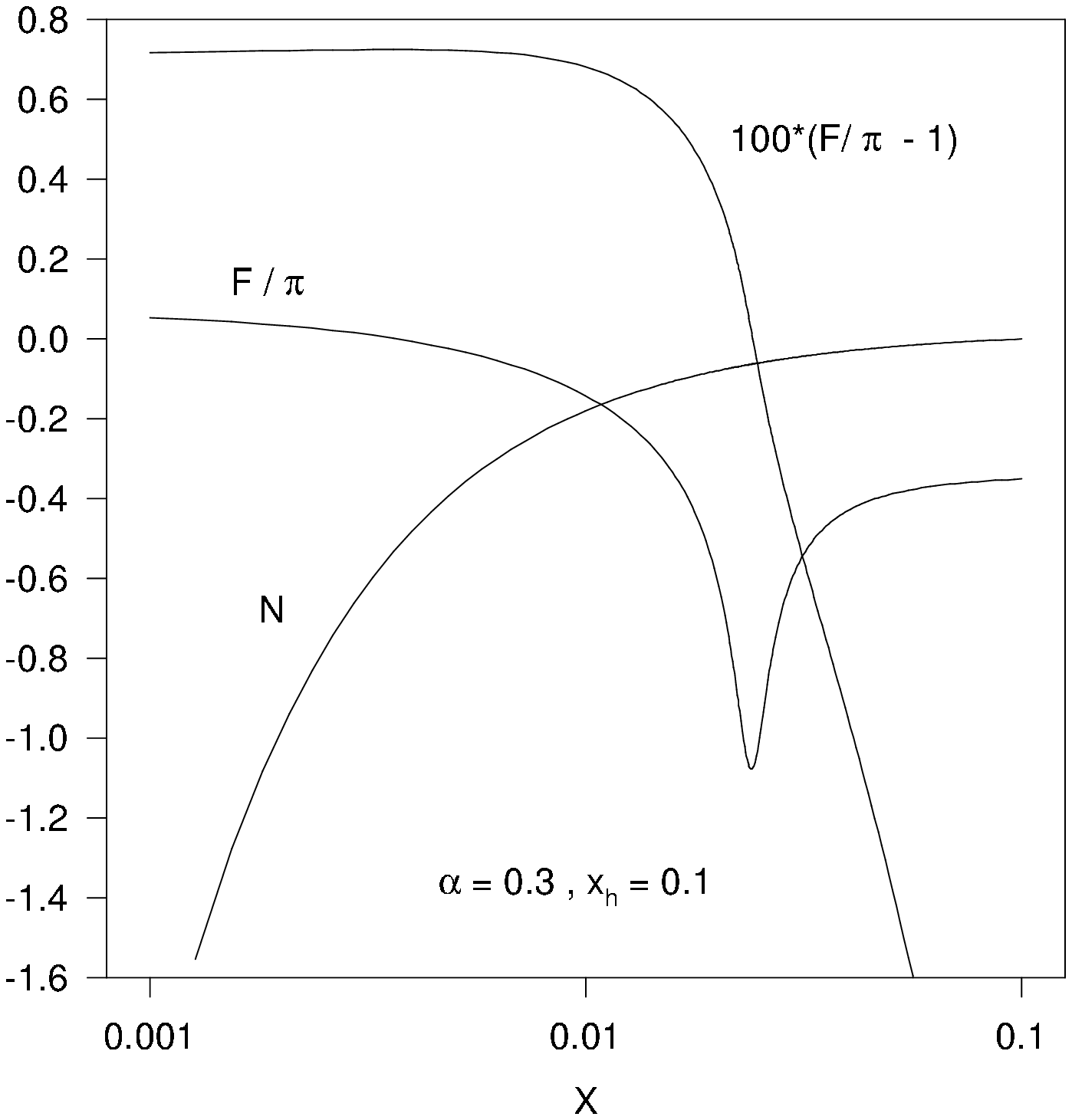}
\vskip -3cm
\caption{\label{fig4}
The profiles, inside the horizon, of the functions $N,f,f'$ 
of the DeSitter-Skyrme
black holes are given
for $\alpha=0.3$, $x_h=0.1$
}
\end{figure}

\end{document}